\documentclass[eqsecnum,aps,superscriptaddress,11pt ]{revtex4}
\usepackage{graphicx}
\usepackage{dcolumn}
\usepackage{xspace}
\usepackage{rotating}

\begin{document}
\title{{\it Ab initio} calculations of `forbidden' transition probabilities and lifetimes of low-lying states in V$^{4+}$ }
\vspace*{0.5cm}

\author{Gopal Dixit$^1$, B.K. Sahoo$^2$, P.C. Deshmukh$^1$, R.K. Chaudhuri$^3$, Sonjoy Majumder$^1$ \\
\vspace{0.3cm}
{\it $^1$Department of Physics, Indian Institute of Technology-Madras, Chennai-600 036, India} \\
$^2$ Max Planck Institute for the Physics of Complex Systems, N\"othnitzer stra{\ss}e 38, D-01187 Dresden,
Germany\\
$^3$ Indian Institute of Astrophysics, Bangalore-34, India}
\date{\today}

\begin{abstract}
\noindent
Electric quadrupole (E2) and magnetic dipole (M1) 
transition amplitudes among the low-lying states of quadruply ionized vanadium V$^{4+}$, important in various field of experimental and astrophysics are
 presented very accurately. Most of these results are reported for the first 
time in the literature. Relativistic coupled-cluster theory with
single, double and leading triple excitations has been
employed for these calculations. Estimation of different correlation effects
arising through the above formalism have been highlighted by studying core and
valence electrons excitations to the excited states. The lifetime of the first excited $D$- state is found to be long. 
\end{abstract}
\maketitle

\section{Introduction}

Electromagnetic `forbidden' transitions, especially for lighter neutral systems and their iso-electronic companions, 
are of immense important in atomic experiments due to precise use of metastable states \cite{biemont}. Some of these transitions 
correspond to relatively longer wavelengths compared to normal allowed transitions of same system provide information about the 
thermal Doppler effects in many physical systems \cite{pearl}. Different astronomical features have only been possible to observe from infrared 
and radio transitions. Many of these forbidden transitions of quadruply ionized vanadium (V$^{4+}$) have been related 
to dominant features in the optical spectra of planetary nebulae and the aurora \cite{papike}. Similar transitions have been 
identified with the so called coronal lines emitted by the Sun \cite{charro}. Under certain circumstances, which 
prevail in astrophysics and low density laboratory tokamak plasmas, electric quadrupole (E2) and magnetic dipole 
(M1) transition lines gain intensity and can be used to infer information about plasma temperature and their 
dynamics \cite{farrag}. The intensities of these transitions allow us to measure the concentration of impurity ions in tokamak 
which originate in the high temperature interior of the discharge \cite{farrag}. From many-body points of view, 
the importance of these results lies in the estimation of the accuracy of the electronic wave function through out 
the radial extent of the atomic systems by comparing the results with experimental measurements \cite{sonjoy1}. Also computed results are only means of estimations for many 
of these transitions  wherever  
experimental measurements are difficult.

Here, we present calculated wavelengths and transition amplitudes of V$^{4+}$ 
involving E2 and M1 radiative 
transition amplitudes which are important in astronomy (as mentioned above), plasma research and can be used in many experiments in atomic 
and solid state physics \cite{schlenker,singh,hecht}. The study of forbidden transitions between the fine structure states of the
low-lying $D$-states needs special attantion as they play important role in the doping of impurity in Al$_2$O$_3$ crystal which is used to study high-frequency 
acoustic phonon in crystal \cite{renk}. It is also a good candidate to study electron spin resonance \cite{biasi} and 
electron paramagnetic resonance \cite{graces} in quartz material. The detail knowledge of the resonant core relaxation 
process of V$^{+4}$ \cite{bates} need accurate results of energy levels of this ion and transition amplitudes among them.

One of the most correlation exhaustive many-body approaches, the relativistic coupled cluster method with singles, 
doubles and partially triples (RCCSD(T)) has been employed. This is non-perterbative in nature. 
 We also intend to investigate the core-core and core-valence correlation contributions obtained from the RCCSD(T) method to the M1 and E2
transitions among the low-lying states.

For the one-valence ($v$) open shell system, the exact wave function can be expressed using the RCC approach as
\begin{equation}
|\Psi_v\rangle = e^T\{e^{S_v}\}|\Phi_v\rangle = e^T\{1+S_v\}|\Phi_v\rangle ,
\end{equation}
where the curly brackets represent the normal order form \cite{lindgren} and the reference state
is defined by
\begin{equation}
|\Phi_v\rangle = a_v^{\dagger} |\Phi_{DF}\rangle
\end{equation}
for the closed-shell Dirac-Fock (DF) state, $|\Phi_{DF}\rangle$, of V$^{+5}$.

In the above equation, the $T$ and $S_v$ operators are the RCC excitation operators associate with the closed and open
shell hole-particle excitations, respectively \cite{lindgren,debasish}. The  computationally intensive parts in this
approach is to consider non-linear terms which involve maximum four powers of single ($T_1$) and two powers of double 
($T_2$) normal ordered core-excitation operators and products with valence-virtual single excitation operators. Computational features are found
in the product of $T_1$ and $T_2$ with  $S_{1v}$ and $S_{2v}$, which represent  
single excitation operator from valence orbitals and double excitations from core-valence orbitals, respectively \cite{sahoo04}.
Leading order triple excitations \cite{sahoo04,sahoo06} are obtained from the latter part.
For computational simplicity, the $T$ amplitudes are calculated first for the closed-shell
$V^{5+}$ then the corresponding valence orbitals are attached to calculate the $V^{5+}$ wave functions for $V^{4+}$ system \cite{geetha}.

The DF orbitals of $V^{5+}$ are generated from the universal Gaussian type orbital (GTO) basis functions \cite{rajat} 
using $\alpha_0$ = 0.00825 and $\beta$ = 2.91. Number of DF orbitals for different symmetries used in the RCC calculations 
is based on convergent criteria of core correlation energy. 
There are 12, 10, 10, 9 and 8 active orbitals including all core electrons are considered in the RCCSD(T) calculations 
for $l=$ 0, 1, 2, 3, 4 symmetries, respectively.  Other higher energy orbitals are  considered as inactive. 

The matrix element of any operator $D$ can be expressed using the RCC method as
\begin{eqnarray}
D_{fi}& = & \frac{\langle \Psi_f|D|\Psi_i\rangle}{\langle \Psi_f|\Psi_i\rangle} \nonumber \\
&=& \frac{\langle \Phi_f|\{1+{S_f}^{\dag}\}{e^T}^{\dag}De^T\{1+S_i\}|\Phi_i\rangle}
{\langle \Phi_f|\{1+{S_f}^{\dag}\}{e^T}^{\dag}e^T\{1+S_i\}|\Phi_i\rangle} .
\end{eqnarray}
The one-electron reduced matrix elements of M1 and E2 operators are given by \cite{sahoo06,berestetski}.
\begin{equation}
\langle j_f||q_m^{(M1)}||j_i \rangle = \langle j_f||C_m^{(1)}||j_i \rangle \frac{6}{\alpha k} 
\frac{\kappa_i+\kappa_f}{2} \left[ \int dr j_1(kr)(P_{kf}Q_{ki}+Q_{kf}P_{ki})\right]
\end{equation}
and
\begin{equation}
\langle j_f||q_m^{(E2)}||j_i \rangle = \langle j_f||C_m^{(2)}||j_i \rangle \frac{15}{ k^2}  
\left[ \int dr{j_2(kr)(P_{kf}P_{ki}+Q_{kf}Q_{ki})}+j_3(kr) \frac{j_f-j_i-1}{3} (P_{kf}Q_{ki}+Q_{kf}P_{ki})\right]
\end{equation}
respectively, where $j_i$ stand for the total angular momentum and $\kappa_i== \pm \left(j_{i}+\frac{1}{2}\right)$ is 
relativistic angular momentum
quantum numbers of the $i^{th}$ electron orbital. 
The quantity $C_m^{(1)}$ is the Racah tensor \cite{sahoo04} and $j_l(kr)$ is the spherical Bessel function of order $l$. 
$P_{ki}$ and $Q_{ki}$ are the large and small radial components of the Dirac wave functions.

The emission transition rate (in $sec^{(-1)}$) for the E1, E2 and M1 channels from state {\it f} to {\it i} are given by,
\begin{equation}
A^{E1}_{f\rightarrow i} = \frac{2.0261\times10^{18}}{\lambda^{3}[j_f]}S^{E1}_{f\rightarrow i} 
\end{equation}
\begin{equation}
A^{E2}_{f\rightarrow i} = \frac{1.11995\times10^{18}}{\lambda^{5}[j_f]}S^{E2}_{f\rightarrow i}
\end{equation}
\begin{equation} 
A^{M1}_{f\rightarrow i} = \frac{2.69735\times10^{13}}{\lambda^{3}[j_f]}S^{M1}_{f\rightarrow i} ,
\end{equation}
where $[j_f] = 2j_f+1$ is the degeneracy of the $ f$-state, $S$ is the square of the transition amplitude of 
the transition operator $D$, and $\lambda$ (in \AA) are the corresponding transition wavelength.

The lifetime (in sec) of a particular state is the reciprocal of total transition probabilities arising from 
all possible electromagnetic spontaneous transitions to the lower energy levels \cite{sahoo04}.
In Paper-I, we have found excellent agreement between the experimental and computed ionization energies using 
the RCCSD(T) approach. The good agreement between length and velocity form electric 
dipole transition matrix elements in the above paper demonstrates the accuracy of our numerical approach.           
Lifetime of excited ststes have been obtained from  these forbidden transition amplitudes reported here and allowed
transition probabilities obtained from the same RCCSD(T) calculation as Paper-I. 
Table I, shows that the $3d_{5/2}$ state has long lifetime and it can be used as a potential metastable states required
for many atomic experiments. \\
  Ali and Kim \cite{ali} had calculated M1 and E2 transition probabilities between
$3d$ and $4s$ states using Dirac-Fock single-configuration (DFSC) approximation.
We have made comparison with them to our calculated  results in Table II.
The difference between the results are due to the inclusion of electron 
correlation effects through the RCCSD(T) method. It is clear from this table that
the magnetic dipole matrix element between $3d_{3/2}$ and $4s$ was highly underestimated in the DFSC calculation.

\begin{table}[h]
\caption{ Lifetime of the low-lying states in $V^{4+}$}
\begin{tabular}{lr}
\hline
States &               Lifetimes (in sec.)\\ \hline
$3d_{5/2}$      &   3.84E+02\\
$4s_{1/2}$       &  4.55E-05\\
$4p_{1/2}$      &   1.99E-10\\
$4p_{3/2}$      &   1.97E-10\\
$4d_{3/2}$       &  3.77E-10\\
$4d_{5/2}$       &  3.01E-10\\
$4f_{5/2}$       &   1.20E-10\\
$4f_{7/2}$       &   6.97E-11\\
\hline
\label{tab:front1}
\end{tabular}
\end{table}

\begin{table}[h]
\caption{Comparison of $3d\rightarrow 4s$ transition probabilities with DFSC calculations \cite{ali}}.
\begin{tabular}{lrrrr}
\hline
Transitions         &  $A_{M1}$(DFSC)  & $A_{M1}$ (RCC)    &  $A_{E2}$ (DFSC)  & $A_{E2}$(RCC)  \\ \hline
$3d \ ^2D_{3/2} \rightarrow 4s \ ^2S_{1/2}$  &  4.21(-06)    &  2.19(-02)         &  8.56(+03)      & 4.42(+03)   \\
$3d \ ^2D_{5/2} \rightarrow 4s \ ^2S_{1/2}$  &               &                    &  1.27(+04)      & 6.60(+03)   \\
\hline
\label{tab:front2}
\end{tabular}
\end{table}

The effect of the unbound orbitals in the correlation calculation of  
the E2 and M1 transition probabilities in the framework of the RCC approach is studied quantitatively in Tables III and IV, 
respectively, which is reported  first time 
in the literature to our knowledge. We have considered transitions involving  excited states few close to and few comparatively away from
the ground state. As expected, the effect is more on the higher excited states compared to the excited states those are closed to the 
ground state. Point to note that effect is more in E2 transition amplitudes and that can be explained by the dependance of these 
transition amplitudes in the more diffused region compare to M1 transition amplitudes. 

\begin{table}[h]
\caption{ Effect of unbound orbitals of $V^{4+}$ on electric quadrupole  transition amplitude}
\begin{tabular}{llrrrr}
\hline
Terms     &     &    & \multicolumn{2}{c}{Trans. Amplitudes } \\ \hline
          &     &                                              & (with bound orb.) & (with all orb.) \\
$4s_{1/2}$&$\rightarrow 4d_{3/2}$&         & -5.8530           & -5.7383 \\
          &$\rightarrow 4d_{5/2}$&         &  7.1673           & 7.0272 \\
$4d_{3/2}$&$\rightarrow 4d_{5/2}$&        &  -6.1508          & -5.9673  \\
$6s_{1/2}$&$\rightarrow 4d_{3/2}$&        &  0.7921           & 0.7101  \\
          &$\rightarrow 4d_{5/2}$&        &  -0.9823          & -0.8820  \\
$4p_{1/2}$&$\rightarrow 4p_{3/2}$&        & -6.2963           & -6.1510  \\
          &$\rightarrow 6p_{3/2}$&        & -0.1036           & -0.2787  \\
$4p_{3/2}$&$\rightarrow 6p_{3/2}$&        & -0.0327           & -0.1949  \\
          &$\rightarrow 6p_{1/2}$&        & -0.0747           & -0.2562  \\
$6p_{1/2}$&$\rightarrow 6p_{3/2}$&       & -0.5787           & -4.6953  \\
\hline
\label{tab:front3}
\end{tabular}
\end{table}

\begin{table}[h]
\caption{Effect of unbound orbitals of $V^{4+}$ on magnetic dipole transition amplitude}
\begin{tabular}{llrrrr}
\hline
Terms     &     &    & \multicolumn{2}{c}{Transition Amplitude}  \\ \hline
          &     &                         & (with bound orb.) & (with all orb.) \\
$4s_{1/2}$&$\rightarrow 6s_{1/2}$&    & -0.0126  & -0.0227  \\
$4d_{3/2}$&$\rightarrow 4d_{5/2}$&    &  -1.5485  &  -1.5455  \\
$4p_{1/2}$&$\rightarrow 4p_{3/2}$&    &  -1.1544  &  -1.1535   \\
          &$\rightarrow 6p_{1/2}$&    &     -0.0102   &   -0.0116  \\
          &$\rightarrow 6p_{3/2}$&    &  0.0071      &    0.0072   \\
$4p_{3/2}$&$\rightarrow 6p_{1/2}$&    & -0.0082      &    -0.0055  \\
          &$\rightarrow 6p_{3/2}$&    &  -0.0746     &    -0.0820   \\
$6p_{1/2}$&$\rightarrow 6p_{3/2}$&    & -1.5398 &   -1.5398   \\
\hline
\label{tab:front3}
\end{tabular}
\end{table}

Tables V and VI give the magnetic dipole and electric quadrupole transition amplitudes, respectively, 
for most of the low-lying states. They are all relevant astrophysically. The important transitions among these 
from the physics point of view are forbidden transitions among fine-structure states of $3d$ and $4p$. Former one falls in the infrared region,
which has many applications in plasma research and infrared laser spectroscopy \cite{thogersen} and latter one 
falls in the optical region, has immense prospect in different atomic physics experiments. 
We have not reported wavelength comparison for most of other fine structure transitions  fall
far beyond the infrared region. The relatively large differences in the  wavelengths between some of the transitions from
$4f$ states call for further experimental and theoretical investigations on these states.  

Quantitative contributions from different correlation terms for few  M1 and E2 transitions among low laying states  are
presented in tables VII, VIII and IX.  These tables shows comparative estimations of core polarisation, core correlation and pair correlation for these transitions. 
 These tables show  that core-polarization effects ($\overline{D} S_{2i}$) coming from the initial state for all the transitions are negligible. 
Wheras the combined effects of $S_2$ operators of both initial and final states contributing significant for both M1 and E2 transitions amplitudes.

\begin{table}[h]
\caption{Transition wavelengths and transition amplitudes of $V^{4+}$ for magnetic dipole transitions.}
\begin{tabular}{llrrr}
\hline
Transitions     &     &  $\lambda_{NIST}$(\AA) & $\lambda_{RCC}$(\AA)  & Transition amplitudes\\ \hline
$3d_{3/2}$&$\rightarrow 3d_{5/2}$&           &            & -1.5398 \\
          &$\rightarrow 4d_{3/2}$& 340.24    & 340.22     &  0.0684  \\
          &$\rightarrow 4d_{5/2}$& 340.08     & 340.04      &  0.0022  \\
          &$\rightarrow 5d_{3/2}$& 257.74     & 257.58      &  -0.0362  \\
          &$\rightarrow 5d_{5/2}$& 257.70     & 257.53      &  -0.0006  \\
          &$\rightarrow 6d_{3/2}$& 230.25     & 230.08      &  -0.0219  \\
          &$\rightarrow 6d_{5/2}$& 230.23     & 230.06      & -0.0001  \\
          &$\rightarrow 4s_{1/2}$& 675.02     &684.25       &  0.0010  \\
          &$\rightarrow 5s_{1/2}$& 304.67     &306.30       & -0.0004  \\
          &$\rightarrow 6s_{1/2}$& 247.61     &248.25       & -0.0002  \\
$3d_{5/2}$&$\rightarrow 4d_{3/2}$& 340.97      & 341.04     &  -0.0062  \\
          &$\rightarrow 4d_{5/2}$& 340.80      & 340.86     &  0.1860  \\
          &$\rightarrow 5d_{3/2}$& 258.16      & 258.05     &  0.0027  \\
          &$\rightarrow 5d_{5/2}$& 258.11      & 258.00     &  -0.0984  \\
          &$\rightarrow 6d_{3/2}$& 230.58      & 230.45     &  0.0013 \\
          &$\rightarrow 6d_{5/2}$& 230.56      & 230.44     &  -0.0598  \\
$4d_{3/2}$&$\rightarrow 4d_{5/2}$&             &            &  -1.5455  \\
          &$\rightarrow 5d_{3/2}$& 1062.99      & 1063.24    &  -0.1248  \\
          &$\rightarrow 5d_{5/2}$& 1062.23      & 1062.49     &  -0.0015  \\
          &$\rightarrow 6d_{3/2}$& 712.24      & 712.36        & -0.0540 \\
          &$\rightarrow 6d_{5/2}$& 712.05      & 712.17        & 0.0008 \\
$4d_{5/2}$&$\rightarrow 5d_{3/2}$& 1064.62      & 1062.18     &  0.0050  \\
          &$\rightarrow 5d_{5/2}$& 1063.86      & 1061.36     &  -0.3485  \\
          &$\rightarrow 6d_{3/2}$& 712.97     & 711.53    &  0.0006  \\
          &$\rightarrow 6d_{5/2}$& 712.79     & 711.38    &  -0.1508  \\
$5d_{3/2}$&$\rightarrow 5d_{5/2}$&           &        & -1.5436\\
          &$\rightarrow 6d_{3/2}$& 2158.58      & 2155.39       &0.1210\\
          &$\rightarrow 6d_{5/2}$& 2142.40     & 2153.97      &0.0012\\
$5d_{5/2}$&$\rightarrow 6d_{3/2}$& 2161.69     &2158.76       & -0.0044\\
          &$\rightarrow 6d_{5/2}$& 2159.95      &2157.34        & -0.0044\\
$6d_{3/2}$&$\rightarrow 6d_{5/2}$&            &                &-1.5453\\
$4s_{1/2}$&$\rightarrow 5s_{1/2}$& 555.32       &554.40         &-0.5404\\
          &$\rightarrow 6s_{1/2}$& 391.06       &389.52         &-0.0227\\
$5s_{1/2}$&$\rightarrow 6s_{1/2}$& 1322.08      &1309.81        &0.0496\\
$4p_{1/2}$&$\rightarrow 4p_{3/2}$& 78971.47     &77464.73       &-1.1535\\
          &$\rightarrow 5p_{1/2}$& 689.14       &688.02         &-0.0207\\
          &$\rightarrow 5p_{3/2}$& 689.69       &685.73         &0.0053\\
          &$\rightarrow 6p_{1/2}$& 478.40       &478.18         &-0.0116\\
          &$\rightarrow 6p_{3/2}$& 477.82       &478.23         &0.0072\\
\hline
\label{tab:front3}
\end{tabular}
\end{table}

\begin{table}[h]
\begin{tabular}{llrrr}
\hline
Transitions     &     &  $\lambda_{NIST}$(\AA) & $\lambda_{RCC}$(\AA)  & Transition amplitudes \\ \hline
$4p_{3/2}$&$\rightarrow 5p_{1/2}$& 695.21       &694.18         &0.0072\\
          &$\rightarrow 5p_{3/2}$& 692.72       &691.85         &-0.1305\\
          &$\rightarrow 6p_{1/2}$& 481.32       &481.15         &-0.0055\\
          &$\rightarrow 6p_{3/2}$& 480.73       &481.20         &-0.0820\\
$5p_{1/2}$&$\rightarrow 5p_{3/2}$&              &               &-1.1530\\
          &$\rightarrow 6p_{1/2}$& 1564.46      &1567.89        &0.0179\\
          &$\rightarrow 6p_{3/2}$& 1558.23      &1568.41        &-0.0077\\
$5p_{3/2}$&$\rightarrow 6p_{1/2}$& 1577.24      &1579.91        &0.0061\\
          &$\rightarrow 6p_{3/2}$& 1570.91      &1580.45        &0.1222\\
$6p_{1/2}$&$\rightarrow 6p_{3/2}$&              &               &-1.4409\\
$4f_{5/2}$&$\rightarrow 4f_{7/2}$&              &               &1.8435\\
$5g_{7/2}$&$\rightarrow 5g_{9/2}$&              &               &-2.1081\\
          &$\rightarrow 6g_{7/2}$& 2970.51      &2850.75        &0.0379\\
          &$\rightarrow 6g_{9/2}$& 2970.45      &2851.10        &-0.0001\\
$5g_{9/2}$&$\rightarrow 6g_{7/2}$& 2970.64      &2850.29        &0.0000\\
          &$\rightarrow 6g_{9/2}$& 2970.58      &2850.64        &0.0663\\
$6g_{7/2}$&$\rightarrow 6g_{9/2}$&              &               &-2.1080\\
\hline
\label{tab:front3}
\end{tabular}
\end{table}

\begin{table}[h]
\caption{Transition wavelengths and transition amplitudes of $V^{4+}$ for electric quadrupole transitions.}
\begin{tabular}{llrrr}
\hline
Transitions     &     &  $\lambda_{NIST}$(\AA) & $\lambda_{RCCSD(T)}$(\AA)  & Transition amplitudes \\ \hline
$3d_{3/2}$&$\rightarrow 3d_{5/2}$&               &                 &-0.7475\\
          &$\rightarrow 4d_{3/2}$& 340.24        &340.22           &1.0870\\
          &$\rightarrow 4d_{5/2}$& 340.08        &341.71           &0.7290\\
          &$\rightarrow 5d_{3/2}$& 257.74        &258.63           &-0.3948\\
          &$\rightarrow 5d_{5/2}$& 257.70        &258.53           &-0.2659\\
          &$\rightarrow 6d_{3/2}$& 230.25        &230.95           &-0.1950\\
          &$\rightarrow 6d_{5/2}$& 230.23        &230.93           &-0.1321\\
          &$\rightarrow 4s_{1/2}$& 675.02        &684.25           &-1.4876\\
          &$\rightarrow 5s_{1/2}$& 304.67        &306.30           &0.0683\\
          &$\rightarrow 6s_{1/2}$& 247.61        &248.25           &0.0269\\
          &$\rightarrow 5g_{7/2}$& 240.17        &240.98           &-0.8463\\
          &$\rightarrow 6g_{7/2}$& 222.21        &222.19           &-0.6366\\
$3d_{5/2}$&$\rightarrow 4d_{3/2}$& 340.97        &341.04           &-0.7357\\
          &$\rightarrow 4d_{5/2}$& 340.80        &340.86           &1.4290\\
          &$\rightarrow 5d_{3/2}$& 258.16        &258.05           &0.2682\\
          &$\rightarrow 5d_{5/2}$& 258.11        &258.00           &-0.5192\\
          &$\rightarrow 6d_{3/2}$& 230.58        &230.45           &0.1337\\
          &$\rightarrow 6d_{5/2}$& 230.56        &230.44           &-0.2574\\
          &$\rightarrow 4s_{1/2}$& 677.88        &680.17           &-1.8310\\
          &$\rightarrow 5s_{1/2}$& 305.25        &305.44           &0.0829\\
          &$\rightarrow 6s_{1/2}$& 247.99        &247.68           &0.0329\\
          &$\rightarrow 5g_{7/2}$& 240.53        &241.39           &0.2839\\
          &$\rightarrow 5g_{9/2}$& 240.53        &241.39           &-1.0037\\
          &$\rightarrow 6g_{7/2}$& 222.51        &222.54           &0.2133\\
          &$\rightarrow 6g_{9/2}$& 222.51        &222.55           &-0.7539\\
$4d_{3/2}$&$\rightarrow 4d_{5/2}$&               &                 &-5.9673\\
          &$\rightarrow 5d_{3/2}$& 1062.99       &1060.41          &-5.5444\\
          &$\rightarrow 5d_{5/2}$& 1062.23       &1059.60          &-3.8912\\
          &$\rightarrow 6d_{3/2}$& 712.24        &710.74           &-1.8378\\
          &$\rightarrow 6d_{5/2}$& 712.05        &710.59           &-1.3032\\
          &$\rightarrow 4s_{1/2}$& 686.06        &684.00           &5.7383\\
          &$\rightarrow 5s_{1/2}$& 2914.22       &2925.88          &9.4581\\
          &$\rightarrow 6s_{1/2}$& 909.48        &904.77           &-0.7101\\
          &$\rightarrow 5g_{7/2}$& 816.61        &826.14           &16.3355\\
          &$\rightarrow 6g_{7/2}$& 640.52        &640.52           &6.1353\\
\hline
\label{tab:front3}
\end{tabular}
\end{table}

\begin{table}[h]
\begin{tabular}{llrrr}
\hline
Transitions     &     &  $\lambda_{NIST}$(\AA) & $\lambda_{RCCSD(T)}$(\AA)  & Transition amplitudes\\ \hline
$4d_{5/2}$&$\rightarrow 5d_{3/2}$& 1064.62       &1062.18          &3.9140\\
          &$\rightarrow 5d_{5/2}$& 1063.86       &1061.36          &-7.2728\\
          &$\rightarrow 6d_{3/2}$& 712.97        &711.53           &1.3062\\
          &$\rightarrow 6d_{5/2}$& 712.79        &711.38           &-2.4142\\
          &$\rightarrow 4s_{1/2}$& 685.38        &683.27           &7.0272\\
          &$\rightarrow 5s_{1/2}$& 2926.53       &2939.35          &11.6086\\
          &$\rightarrow 6s_{1/2}$& 910.68        &906.06           &-0.8820\\
          &$\rightarrow 5g_{7/2}$& 817.57        &827.21           &-5.4566\\
          &$\rightarrow 5g_{9/2}$& 817.56        &827.25           &19.2923\\
          &$\rightarrow 6g_{7/2}$& 641.11        &641.16           &-2.0469\\
          &$\rightarrow 6g_{9/2}$& 641.12        &641.18           &7.2372\\
$5d_{3/2}$&$\rightarrow 5d_{5/2}$&               &                 &-19.0483\\
          &$\rightarrow 6d_{3/2}$& 2158.58       &2155.39          &15.9663\\
          &$\rightarrow 6d_{5/2}$& 2156.85       &2153.97          &11.2383\\
          &$\rightarrow 4s_{1/2}$& 416.95        &415.80           &0.1076\\
          &$\rightarrow 5s_{1/2}$& 1673.36       &1663.21          &18.2095\\
          &$\rightarrow 6s_{1/2}$& 6298.00       &6164.49          &-29.2917\\
          &$\rightarrow 5g_{7/2}$& 3523.20       &3739.47          &44.8003\\
          &$\rightarrow 6g_{7/2}$& 1611.66       &1617.59          &-27.9356\\
$5d_{5/2}$&$\rightarrow 6d_{3/2}$& 2161.69       &2158.76          &-11.3019\\
          &$\rightarrow 6d_{5/2}$& 2159.95       &2157.34          &20.9536\\
          &$\rightarrow 4s_{1/2}$& 416.83        &415.67           &0.1240\\
          &$\rightarrow 5s_{1/2}$& 1671.50       &1661.21          &22.2903\\
          &$\rightarrow 6s_{1/2}$& 6324.53       &6192.12          &-35.9371\\
          &$\rightarrow 5g_{7/2}$& 3531.49       &3749.62          &-14.9385\\
          &$\rightarrow 5g_{9/2}$& 3531.31       &3770.43          &52.8170\\
          &$\rightarrow 6g_{7/2}$& 1613.40       &1619.49          &9.3478\\
          &$\rightarrow 6g_{9/2}$& 1613.38       &1619.60          &-33.0483\\
$6d_{3/2}$&$\rightarrow 6d_{5/2}$&               &                 &-49.7139\\
          &$\rightarrow 4s_{1/2}$& 349.45        &348.55           &0.1295\\
          &$\rightarrow 5s_{1/2}$& 942.62        &938.79           &0.5210\\
          &$\rightarrow 6s_{1/2}$& 3284.21       &3314.19          &47.8467\\
          &$\rightarrow 5g_{7/2}$& 5573.05       &5088.16          &-29.2475\\
          &$\rightarrow 6g_{7/2}$& 6361.01       &6483.01          &128.4251\\
$6d_{5/2}$&$\rightarrow 4s_{1/2}$& 349.40        &348.52           &0.1553\\
          &$\rightarrow 5s_{1/2}$& 942.29        &938.52           &0.6557\\
          &$\rightarrow 6s_{1/2}$& 3280.21       &3310.84          &58.5607\\
          &$\rightarrow 5g_{7/2}$& 5561.54       &5080.26          &9.7363\\
          &$\rightarrow 5g_{9/2}$& 5562.00       &5078.78          &-34.4251\\
          &$\rightarrow 6g_{7/2}$& 6376.81       &6495.88          &-42.8137\\
          &$\rightarrow 6g_{9/2}$& 6375.81       &6497.70          &151.3704\\
\hline
\label{tab:front3}
\end{tabular}
\end{table}
\begin{table}[h]
\begin{tabular}{llrrr}
\hline
Transitions     &     &  $\lambda_{NIST}$(\AA) & $\lambda_{RCCSD(T)}$(\AA)  & Transition amplitudes \\ \hline
$5g_{7/2}$&$\rightarrow 5g_{9/2}$&               &                 &-14.0132\\
          &$\rightarrow 6g_{7/2}$& 2970.51       &2850.75          &25.5408\\
          &$\rightarrow 6g_{9/2}$& 2970.45       &2851.10          &7.7503\\
$5g_{9/2}$&$\rightarrow 6g_{7/2}$& 2970.64       &2850.29          &-7.7543\\
          &$\rightarrow 6g_{9/2}$& 2970.58       &2850.64          &28.8149\\
$6g_{7/2}$&$\rightarrow 6g_{9/2}$&               &                 &-44.2007\\
$4p_{1/2}$&$\rightarrow 4p_{3/2}$& 78971.47      &77464.73         &-6.1510\\
          &$\rightarrow 5p_{3/2}$& 686.69        &685.73           &-3.5604\\
          &$\rightarrow 6p_{3/2}$& 477.82        &478.23           &-0.2787\\
          &$\rightarrow 4f_{5/2}$& 697.92        &709.04           &8.4251\\
$5p_{1/2}$&$\rightarrow 4p_{3/2}$& 695.21        &694.18           &-3.6670\\
          &$\rightarrow 5p_{3/2}$&               &                 &-21.5598\\
          &$\rightarrow 6p_{3/2}$& 1558.23       &1568.41          &2.9529\\
          &$\rightarrow 4f_{5/2}$& 54796.32      &23207.88         &13.7048\\
$6p_{1/2}$&$\rightarrow 4p_{3/2}$& 481.32        &481.15           &-0.2562\\
          &$\rightarrow 5p_{3/2}$& 1577.24       &1579.91          &3.0066\\
          &$\rightarrow 6p_{3/2}$&               &                 &-4.6953\\
          &$\rightarrow 4f_{7/2}$& 1521.03       &1468.67          &-0.7073\\
$4p_{3/2}$&$\rightarrow 5p_{3/2}$& 692.72        &691.85           &-3.4660\\
          &$\rightarrow 6p_{3/2}$& 480.73        &481.20           &-0.1949\\
          &$\rightarrow 4f_{5/2}$& 704.14        &715.59           &-4.5351\\
          &$\rightarrow 4f_{7/2}$& 706.25        &715.71           &-11.0889\\
$5p_{3/2}$&$\rightarrow 6p_{3/2}$& 1570.91       &1580.45          &2.6031\\
          &$\rightarrow 4f_{5/2}$& 37839.08      &20858.40         &-7.3110\\
          &$\rightarrow 4f_{7/2}$& 32616.42      &20755.93         &-17.8582\\
$6p_{3/2}$&$\rightarrow 4f_{5/2}$& 1515.14       &1469.13          &0.3491\\
          &$\rightarrow 4f_{7/2}$& 1505.49       &1468.62          &0.8998\\
$4f_{5/2}$&$\rightarrow 4f_{7/2}$&               &                 &5.5868\\
\hline
\label{tab:front3}
\end{tabular}
\end{table}

\begin{sidewaystable}
\centering
\begin{tabular}{lccccccccc}
\hline
RCC terms & $3d_{3/2}\rightarrow 3d_{5/2}$ & $3d_{3/2}\rightarrow 4d_{3/2}$ & $3d_{3/2}\rightarrow 4s_{1/2}$ & $3d_{3/2}\rightarrow 5s_{1/2}$ & $3d_{3/2}\rightarrow 5g_{7/2}$ & $3d_{5/2}\rightarrow 4d_{3/2}$ & $3d_{5/2}\rightarrow 4s_{1/2}$ & $3d_{5/2}\rightarrow 5s_{1/2}$ & $3d_{5/2}\rightarrow 5g_{7/2}$ \\
\hline
 & & & & & & & & & \\
  Dirac-Fock D & -0.8554 & 1.1077 & -1.5552 & 0.0624 & -0.9137 & -0.7289 & -1.9121 & 0.0748 & 0.3061 \\
$\overline{D}$ & -0.8466 & 1.1079 & -1.5505 & 0.0604 & -0.9125 & -0.7287 & -1.9060 & 0.0722 & 0.3057 \\
$\overline{D} S_{1i}$ & 0.0180 & -0.0896 & 0.0882 & 0.0284 & 0.0774 & 0.0585 & 0.1078 & 0.0348 & -0.0257 \\
$S_{1f}^{\dagger} D$ & 0.0178 & 0.0252 & -0.0019 & -0.0351 & -0.0096 & -0.0355 & -0.0023 & -0.0432 & 0.0032 \\
$\overline{D} S_{2i}$ & 0.0000 & 0.0000 & 0.0000 & 0.0000 & 0.0000 & 0.0000 & 0.0000 & 0.0000 & 0.0000 \\
$S_{2f}^{\dagger} D$ & 0.0625 & 0.0501 & -0.0244 & 0.0092 & -0.0036 & -0.0337 & -0.0318 & 0.0124 & 0.0014 \\
$S_{1f}^{\dagger} \overline{D} S_{1i}$ & 0.0000  & -0.0012 & 0.0000 & 0.0000 & 0.0000 & 0.0000 & 0.0000 & 0.0000 & 0.0000\\
$S_{1f}^{\dagger} \overline{D} S_{2i}$ &  0.0000 & 0.0000 & 0.0000 & 0.0000 & 0.0000 & 0.0000 & 0.0000 & 0.0000 & 0.0000 \\
$S_{2f}^{\dagger} \overline{D} S_{1i}$ & 0.0001 & 0.0010  & -0.0013 & 0.0006 & -0.0004 & -0.0006 & -0.0016 & 0.0007 & 0.0001 \\
$S_{2f}^{\dagger} \overline{D} S_{2i}$ & -0.0078 & 0.0039 & -0.0104 & 0.0048 & -0.0021 & -0.0026 & -0.0127 & 0.0053 & 0.0006 \\
\hline\\
\multicolumn{5}{c}{\textbf{Important effective two-body terms of $\overline{D}$ }} \\
\hline
 & & & & & & & & & \\
$S_{2f}^{\dagger} D T_1$ & 0.0003 & 0.0002 & 0.0002 & -0.0001 & 0.0000 & -0.0002 & 0.0002 & -0.0001 & 0.0000 \\
$T_{1}^{\dagger} D S_{2i}$ & 0.0003 & 0.0000 & 0.0001 & 0.0000 & 0.0000 & 0.0000 & 0.0002 & -0.0001 & 0.0000 \\
$T_{2}^{\dagger} D S_{2i}$ &  0.0000 & 0.0000 & 0.0000 & 0.0000 & 0.0000 & 0.0000 & 0.0000 & 0.0000 & 0.0000 \\
$S_{2f}^{\dagger} D T_2$ &  0.0000 & 0.0000 & 0.0000 & 0.0000 & 0.0000 & 0.0000 & -0.0026 & 0.0015 & 0.0000 \\
 Norm. & 0.0075 & -0.0107 & 0.0143 & -0.0004 & 0.0044 & 0.0072 & 0.0175 & -0.0005 & -0.0014 \\
\hline
\hline
\end{tabular}
\caption{Explicit contributions from the RCCSD(T) calculations to the absolute magnitude of reduced E2 transitions matrix elements in a.u.}
\label{tab:front3}
\end{sidewaystable}

\begin{sidewaystable}
\begin{tabular}{lccccccccc}
\hline
\hline
RCC terms & $3d_{5/2}\rightarrow 5g_{9/2}$ & $4d_{3/2}\rightarrow 4s_{1/2}$ & $4d_{3/2}\rightarrow 5s_{1/2}$ & $4d_{3/2}\rightarrow 5g_{7/2}$ & $4p_{1/2}\rightarrow 4p_{3/2}$ & $4p_{1/2}\rightarrow 4f_{5/2}$ & $4p_{3/2}\rightarrow 4f_{5/2}$ & $4p_{3/2}\rightarrow 4f_{7/2}$ & $4f_{5/2}\rightarrow 5f_{7/2}$ \\
\hline
 & & & & & & & & & \\
  Dirac-Fock D & -1.0825 & 5.9612 & 10.0120 & 17.4676 & -6.4365 & 9.1096 & -4.9059 & -12.0170 & 6.5586 \\
$\overline{D}$ & -1.0810 & 5.9612 & 10.0120 & 17.4678 & -6.4367 & 9.1094 & -4.9061 & -12.0174 & 6.5585 \\
$\overline{D} S_{1i}$ & 0.0912 & 0.0177 &-0.6124 & -1.0685 & 0.1011 & -0.2782 & 0.1485 & 0.3639 & -0.2998 \\
$S_{1f}^{\dagger} D$ & -0.0110 & -0.1701 & 0.1189 & 0.0334 & 0.0982 & -0.0815 & 0.0463 & 0.1145 &-0.3017 \\
$\overline{D} S_{2i}$ & 0.0000 & 0.0000 & 0.0000 & 0.0000 & 0.0000 & 0.0000 & 0.0000 & 0.0000 & 0.0000 \\
$S_{2f}^{\dagger} D$ & -0.0047 & -0.0430 & 0.0060 & -0.0264 & 0.0654 & -0.0557 & 0.0302 & 0.0733 & -0.0195 \\
$S_{1f}^{\dagger} \overline{D} S_{1i}$ & 0.0000  & 0.0000 & 0.0000 & 0.0000 & 0.0000 & 0.0000 & 0.0000 & 0.0000 & 0.0000\\
$S_{1f}^{\dagger} \overline{D} S_{2i}$ &  0.0000 & 0.0000 & 0.0000 & 0.0000 & 0.0000 & 0.0000 & 0.0000 & 0.0000 & 0.0000 \\
$S_{2f}^{\dagger} \overline{D} S_{1i}$ &-0.0005 &-0.0010  & 0.0010 & 0.0001 & 0.0007 & -0.0002 & 0.0001 & 0.0003 &-0.0011 \\
$S_{2f}^{\dagger} \overline{D} S_{2i}$ & -0.0026 & 0.0277 & -0.0013 & 0.0113 & -0.0316 & 0.0479 & -0.0245 & -0.0614 & 0.0389 \\
\hline\\
\multicolumn{5}{c}{\textbf{Important effective two-body terms of $\overline{D}$ }} \\
\hline
 & & & & & & & & & \\
$S_{2f}^{\dagger} D T_1$ & 0.0000 &-0.0002 & 0.0001 & 0.0000 & 0.0003 & -0.0001 & 0.0000 & 0.0001 & 0.0000 \\
$T_{1}^{\dagger} D S_{2i}$ & 0.0000 &-0.0001 & 0.0001 & 0.0000 & 0.0000 &-0.0003 & 0.0001 & 0.0003 & 0.0000 \\
$T_{2}^{\dagger} D S_{2i}$ &  0.0000 & 0.0000 & 0.0000 & 0.0000 & 0.0055 & 0.0029 & -0.0017 & -0.0040 & 0.0000 \\
$S_{2f}^{\dagger} D T_2$ &  0.0000 & 0.0003 & -0.0002 & 0.0000 & -0.0025 & 0.0000 & 0.0000 & 0.0000 & 0.0000 \\
 Norm. & 0.0051 & -0.0543 & -0.0657 & -0.0822 & 0.0557 & -0.3186 & 0.1714 & 0.4418 & -0.3881 \\
\hline
\hline
\end{tabular}
\caption{Explicit contributions from the RCCSD(T) calculations to the absolute magnitude of reduced E2 transitions matrix elements in a.u.}
\label{tab:front3}
\end{sidewaystable}

\begin{table}[h]
\begin{tabular}{lrrrrrr}
\hline 
RCC terms & $3d_{3/2}\rightarrow 3d_{5/2}$ & $3d_{3/2}\rightarrow 4d_{3/2}$ & $4s_{1/2}\rightarrow 5s_{1/2}$ & $4p_{1/2}\rightarrow 4p_{3/2}$ & $4f_{5/2}\rightarrow 4f_{7/2}$ & $5g_{7/2}\rightarrow 5g_{9/2}$ \\
\hline
 & & & & & & \\
  Dirac-Fock D & -1.5489 & -0.0004 & 0.0004 & -1.1545 & 1.8515 & -2.1081 \\
$\overline{D}$ & -1.5291 & -0.0020 & 0.0015 & -1.1538 & 1.8515 & -2.1081 \\
$\overline{D} S_{1i}$ & 0.0001 & 0.0511 &-0.0440 & 0.0002 & 0.0000 & 0.0000 \\
$S_{1f}^{\dagger} D$ & 0.0000 & 0.0000 & 0.0000 & -0.0002 & 0.0000 & 0.0000 \\
$\overline{D} S_{2i}$ & 0.0000 & 0.0000 & 0.0000 & 0.0000 & 0.0000 & 0.0000 \\
$S_{2f}^{\dagger} \overline{D}$ & 0.0000 & 0.0000 & 0.0000 & 0.0000 & 0.0000 & 0.0000 \\
$S_{1f}^{\dagger} \overline{D} S_{1i}$ & 0.0000  & 0.0017 & -0.0006 & 0.0000 & 0.0000 & 0.0000 \\
$S_{1f}^{\dagger} \overline{D} S_{2i}$ &  0.0000 & 0.0000 & 0.0000 & 0.0000 & 0.0000 & 0.0000 \\
$S_{2f}^{\dagger} \overline{D} S_{1i}$ & 0.0000 & 0.0000  & 0.0000 & 0.0000 & 0.0000 & 0.0000 \\
$S_{2f}^{\dagger} \overline{D} S_{2i}$ & -0.0187 & 0.0199 & -0.0118 & -0.0106 & 0.1203 & -0.0008 \\
\hline\\
\multicolumn{5}{c}{\textbf{Important effective two-body terms of $\overline{D}$ }} \\
\hline
 & & & & & &  \\
$S_{2f}^{\dagger} D T_1$ & 0.0011 &-0.0009 & 0.0002 & 0.0001 & 0.0000 & 0.0000 \\
$T_{1}^{\dagger} D S_{2i}$ & 0.0011 &-0.0006 & 0.0001 & 0.0001 & 0.0000 & 0.0000 \\
$T_{2}^{\dagger} D S_{2i}$ &  0.0000 & 0.0000 & 0.0000 & 0.0000 & 0.0000 & 0.0000 \\
$S_{2f}^{\dagger} D T_2$ &  0.0000 & 0.0000 & 0.0000 & 0.0000 & 0.0000 & 0.0000 \\
 Norm. & 0.0154 & -0.0006 & 0.0003 & 0.0104 & -0.1280 & 0.0008 \\
\hline
\hline
\end{tabular}
\caption{Explicit contributions from the RCCSD(T) calculations to the absolute magnitude of reduced M1 transitions matrix elements in a.u.}
\label{tab:front3}
\end{table} 

\section{conclusion}
Lifetimes of the low-lying bound states of V$^{+4}$ have been calculated using highly correlated relativistic coupled-cluster
approach. Long lifetimes have been observed for the first excited $D$- states and they can be used as potential metastable state for experiments in physics. 
Magnetic dipole and electric quadrupole transition amplitudes
among bound states of the system, important for astronomical observations and plasma researches are estimated for the first 
time in the literature for most of the cases. Especially, forbidden transitions between the fine structure $4p$ states
may be considered for different atomic experiments of fundamental physics due to its optical transition line. 
We have also highlighted different correlation effects arising through the RCCSD(T) method.
\section{Acknowledgment}

\end{document}